\documentclass[12pt]{article}

\usepackage{fullpage}
\usepackage{setspace}
\usepackage{graphicx}
\usepackage{epsfig} 
\usepackage{epstopdf}

\usepackage[usenames,dvipsnames]{color}
\usepackage[usenames,dvipsnames,svgnames,table]{xcolor}
\usepackage[normalem]{ulem} 
\usepackage{amsmath}  
\usepackage{latexsym}
\usepackage{multirow} 
\usepackage[super,comma,sort&compress]{natbib}

\usepackage{algorithm}
\usepackage{algorithmic}

\usepackage{arydshln} 
\DeclareMathAlphabet{\mathpzc}{OT1}{pzc}{m}{it}

\usepackage[hypertexnames=false,linktocpage=true]{hyperref}
\hypersetup{colorlinks=true,linkcolor=blue,anchorcolor=blue,citecolor=blue,filecolor=blue,urlcolor=blue,bookmarksnumbered=true,pdfview=FitB}

\definecolor{gray}{gray}{0.8}
\definecolor{darkgray}{gray}{0.6}

\newcommand{\db}{{\mathbf d}}

\newcommand{\mb}{{\mathbf m}}

\newcommand{\pb}{{\mathbf p}}           
\newcommand{\qb}{{\mathbf q}}

\newcommand{\vb}{{\mathbf v}}           
\newcommand{\wb}{{\mathbf w}}

\usepackage{dsfont}

\DeclareMathAlphabet{\mathpzc}{OT1}{pzc}{m}{it}

\newcommand{\uj}{{\bf u_j}}
\newcommand{\R}{{R}}

\title{The Symmetricity of Normal Modes in Symmetric Complexes}
\author{
Guang Song$^{1,2,3*}$ \\
  $^1$Department of Computer Science, \\
  $^2$Program of Bioinformatics and Computational Biology, \\
  $^3$L. H. Baker Center for Bioinformatics and Biological Statistics, \\
  Iowa State University, Ames, IA 50011, USA \\
  \\
  $^*$Correspondence to: Guang Song, Tel: 515-294-1696; \\
  Fax: 515-294-0258; E-mail: gsong@iastate.edu
}

\begin{document}
\maketitle

\begin{abstract}

In this work, we look at the symmetry of normal modes in symmetric structures, particularly structures with cyclic symmetry. 
We show that normal modes of symmetric structures have different levels of symmetry, or symmetricity. One novel theoretical result of this work is that, for a ring structure with $m$ subunits, the symmetricity of the normal modes falls into $m$ groups of equal size, with normal modes in each group having the same symmetricity. 
The normal modes in each group can be computed separately,
using a much smaller amount of memory and time (up to $m^3$ less), thus making it applicable to larger complexes.  
We show that normal modes with perfect symmetry or anti-symmetry have no degeneracy while the rest of the modes have a degeneracy of two.
We show also how symmetry in normal modes correlates with symmetry
in structure. While a broken symmetry in structure generally leads to a loss 
of symmetricity in symmetric normal modes, the symmetricity of some
symmetric normal modes is preserved even when symmetry in structure is broken.
This work suggests a deeper reason for the existence 
of symmetric complexes: that they may be formed not only for structural purpose, 
but likely also for a dynamical reason, that certain symmetry is needed by a given complex to obtain certain symmetric motions that are functionally critical.


%

\end{abstract}

\section{Introduction}
Many proteins in cell need to form structure complexes 
in order to function. 
In E-coli, it was estimated that over 80\% of 
proteins form structural complexes~\cite{Goodsell00}. Most of the
structures reported in PDB form complexes too. According to a recent 
work by Levy et al.~\cite{Levy06}, about 1/2 to 2/3 of the proteins in PDB
are multi-meric, i.e., containing multiple chains. 
Such multi-unit complexes are called oligomers 
or multimers. They represent the quaternary structures of proteins.  
An oligomer is composed of more than one peptide chains, each of 
which is called a subunit of the complex.  
Some of these complexes are composed of monomers of different types, 
which are called heteroligomers, or heteromers, while others are composed of 
identical subunits and are called homoligomers or homomers.  

How proteins form quaternary complexes is an important research topic.
There are a number of work that predict the quaternary structures of 
proteins, such as PQS~\cite{pqs}, PISA~\cite{pisa}, 
3D-complex~\cite{Levy06}, PiQSi~\cite{Levy07}, etc. Some of these 
are fully-automatic~\cite{pqs, pisa}, while others are manually curated~\cite{Levy07}.  
3D-complex~\cite{Levy06} provides a novel hierarchical classification of the organization of protein complexes 
and it uses biological assemblies of PDB as input.
In their recent study~\cite{Teichmann15}, Teichmann and her co-workers delved deeper into 
the organizing principle of protein complexes and found that  
most assembly steps fall into three basic types: ``dimerization,
cyclization, and heteromeric subunit addition''~\cite{Teichmann15}. According to this 
principle, protein complexes can be organized neatly into a ``
periodic table'',~\cite{Teichmann15} which not only provides fresh insights into the 
patterns and formation 
principle of protein complexes but also may be used to predict
complexes that are not yet observed structurally. 

Homomers are more abundant than heteromers. 
One known fact about homomers is that more of them have an  
even number of subunits than odd.~\cite{Levy06} 
Levy et al.~\cite{Levy08} shows that most homomers have either cyclic or dihedral 
symmetry, while other kinds of symmetry are rare. For cyclic structures, Kidera and coworkers~\cite{Kidera12}
found that there is no preference for an even or odd number of subunits, and therefore the reason 
behind the preference for an even number of subunits is that complexes with 
dihedral symmetry always require an even number of subunits~\cite{Kidera12}.  
In summary, most proteins form structure complexes and a majority of these 
complexes are homomers and symmetric. {\em symmetric structures are prevalent.} 
Complexes
with cyclic symmetry are commonly called ring structures. 
In the work, we focus on the normal modes of 
symmetric ring structures. The study can be extended to other symmetric structures and perhaps even pseudo-symmetric structures~\cite{Goodsell00}, of which the subunits are not identical but the overall structure is almost symmetric.   


Normal modes analysis~\cite{Go83,Brooks83,Levitt83}, as a powerful tool for studying protein vibrational
dynamics near equilibrium state,  has been extensively studied and applied 
for the last two decades using coarse-grained models~\cite{Tirion96,bahar97,Hinsen98,Hinsen99,Atilgan01,Tama01,
Keskin02,Li02,XuBahar03,Ma04,
Maragakis05,VanWynsberghe05,Zheng05,
Kurkcuoglu06,Tama06symmetry,Lezon09,
Yang09,Lin10,Kim13,Na14a,Na14b}. Dynamics produced from NMA has provided insightful 
understanding of the functional mechanisms of a wide range of proteins. 
Deeper understandings regarding the properties of normal modes themselves also have 
been obtained, such as the property of low frequency modes, high-frequency modes, 
hot-spot residues~\cite{Ozbek13}, dynamics-residues~\cite{Zheng05},  the universality of the vibrational spectrum~\cite{Hinsen00-ms,Na16a}, 
the effective degeneracy of normal modes~\cite{Na16b}, etc.  
These insightful understandings of normal modes are helpful for 
a more accurate and appropriate use of them in various applications. 

The topic of this study is about another property of 
normal modes, namely, the symmetricity of the normal modes in symmetric structures. 
Since the structures under consideration are symmetric, 
There are a number of interesting questions that can be raised regarding 
their normal modes:
\begin{itemize}
\item One is about {\bf efficiency}. Since 
the complex is symmetric, is there a way to take advantage of the symmetry 
so as to obtain the normal modes 
more efficiently?
\item 
Another is about {\bf degeneracy}. Are all the modes degenerate 
due to the symmetry in structure? What is the level of degeneracy? 
\item 
{\bf Dynamics}. Since the structure is 
symmetric, are the modes/motions also symmetric? How to measure the level 
of a mode's symmetricity? 
\item
Lastly, about {\bf function}. Are the functional 
processes symmetric? If so, how are they related to the intrinsic symmetric modes of symmetric ring structures?
\end{itemize}
In their seminal work, Simonson and Perahia~\cite{Simonson92} addressed the first issue 
regarding efficiency and showed that by using group theory the time for 
computing normal 
modes of symmetric structures can be greatly reduced. Using a similar approach,
Vlijmen and Karplus showed that efficient normal mode analysis could be 
carried out for large systems with icosahedral symmetry~\cite{Vlijmen01} and 
successfully applied it to Icosahedral viruses~\cite{Vlijmen05}.  
In a recent work, Matsunaga et al.~\cite{Kidera12} demonstrated that structure symmetry 
has a determinant effect on the protein dynamics of circularly symmetric structure. They observed that multimers with a
highly composite number of subunits (such as 6 or 12) tend to have 
more inter-unit fluctuations while multimers with a primer number of subunits tend to have more intra-unit fluctuations. 

Our work here focuses on the {\em degeneracy} and {\em dynamics} of symmetric complexes.  
Our previous work showed that 
the functional motions of some symmetric structures are symmetric and are closely
related to symmetric modes~\cite{Na16c}.

\section{Methods}
For symmetric ring structures of $m$ subunits of size $N$ (i.e., $N$ atoms or residues in each subunit), 
the use of symmetry can reduce the size of the matrix that has to be diagonalized
from 3mNx3mN to to a series of smaller matrices of 3Nx3N. 
The most general, rigorous way to take advantage of symmetry 
is to use group theory~\cite{Cotton90,Simonson92, Vlijmen01, Vlijmen05, Kidera12}, which utilizes the character table and irreducible representations 
to find symmetric coordinates~\cite{Cotton90,Simonson92, Vlijmen01, Vlijmen05, Kidera12}. When expressed in symmetric coordinates, the Hessian matrix  
becomes block diagonal.
For the special case of 
symmetric ring structures, the process can be understood using 
a mathematically simpler approach that is based on circulant matrix~\cite{}. 

\subsection{Circulant Matrix}
A circular matrix has the following form~\cite{Davis70,wiki-cm}:
\begin{equation}
\begin{bmatrix}
c_{0}     & c_{1} & \dots  & c_{m-2} & c_{m-1}  \\
c_{m-1} & c_0    & \dots  & c_{m-3}      & c_{m-2}  \\
\vdots  & \vdots &	\vdots    & \vdots  & \vdots   \\
c_{2}  &  c_3      & \dots & c_0  & c_{1}   \\
c_{1}  & c_{2} & \dots  & c_{m-1} & c_0 \\
\end{bmatrix}
\end{equation}
The normalized eigenvectors of a circulant matrix are given by
\begin{equation}
v_j=\frac{1}{\sqrt{m}} (1,~ \omega_j,~ \omega_j^2,~ \ldots,~ \omega_j^{m-1})^T,\quad j=0, 1,\ldots, m-1,
\end{equation}
where $\omega_j=\exp \left(\tfrac{2\pi i j}{m}\right)$ are the $m^{th}$ roots of unity and $i$ is the imaginary unit.

The corresponding eigenvalues are:
\begin{equation}
\lambda_j = c_0+c_{1} \omega_j + c_{2} \omega_j^2 + \ldots + c_{m-1} \omega_j^{m-1}, \qquad j=0, 1,\ldots, m-1.
\end{equation}

\subsection{Small Oscillations of M Identical Masses in a Hoop}
\label{sec:oscillation}

For a spring-mass system of $m$ identical masses connected with $m$ identical springs in a hoop, 
the Kirchhoff matrix is a circulant matrix where $c_0 = 2$, $c_{m-1}=c_1=-1$, and the rest of $c$'s are all zeros. Therefore, the eigenvalues are:
 
\begin{equation}
\lambda_j = c_0+c_{1} \omega_j + c_{m-1} \omega_j^{m-1} = 2 - 2cos(\frac{2\pi j}{m}), \qquad j=0, 1,\ldots, m-1.
\end{equation}

Except for $\lambda_0$ (which is 0) and $\lambda_{\frac{m}{2}}$ when $m$ is even, all the modes are degenerate and have a degeneracy of 2.

\subsection{Symmetric Circulant Block Matrix}
A symmetric circulant block matrix has the following form:
\begin{equation}
H = 
\begin{bmatrix}
A_{0}     & A_{1} & \dots  & A_{m-2} & A_{m-1}  \\
A_{m-1} & A_0    & \dots  & A_{m-3}      & A_{m-2}  \\
\vdots  & \vdots &	\vdots    & \vdots  & \vdots   \\
A_{2}  &  A_3      & \dots & A_0  & A_{1}   \\
A_{1}  & A_{2} & \dots  & A_{m-1} & A_0 \\
\end{bmatrix}
\label{eq:As}
\end{equation}
and $A_k = A'_{m-k}$ since the matrix is symmetric. In the discussion that follows, we limit ourselves to the scenario that $H$ is also real, as is for a Hessian 
matrix. 

An efficient way to obtain $H$'s eigenvalues and eigenvectors exists and was given by Cao~\cite{Cao90}. Part of his proof is included here for clarity. In his proof, Cao constructed a series of $M$ matrices:

\begin{equation}
\tilde{M}_j = \sum_{k=0}^{m-1} \omega_m^{kj}A_k
\end{equation}
where $\omega_m$ is equal to $e^{\frac{2\pi i}{m}}$. 

Specifically, 
\begin{equation}
\tilde{M}_0 = \sum_{k=0}^{m-1} A_k,
\label{eq:M0}
\end{equation}
and when m is even,
\begin{equation}
\tilde{M}_{m/2} = \sum_{k=0}^{m-1} (-1)^k A_k.
\label{eq:anti}
\end{equation}


According to Cao~\cite{Cao90}, define a matrix $F_m$ as:
\begin{equation}
F_m = \frac{1}{\sqrt{m}}
\begin{bmatrix}
1    & 1 & 1 & \dots  & 1  \\
1 & \omega_m    & \omega_m^2 & \dots        & \omega_m^{m-1}  \\
1 & \omega_m^2    & \omega_m^4 & \dots        & \omega_m^{2(m-1)}  \\
\dots  & \dots & \dots    & \dots  & \dots   \\
1 & \omega_m^{m-1}    & \omega_m^{2(m-1)} & \dots        & \omega_m^{(m-1)^2}  \\
\end{bmatrix}
\end{equation}

Assume ${\bf u_j}$ is one of eigenvectors of $\tilde{M}_j$, then 
$\wb$ = ($0_n, 0_n, \dots, {\uj}, \dots, 0_n$)' will be the eigenvectors of 
diag($\tilde{M}_0, \tilde{M}_1, \dots, \tilde{M}_j, \dots, \tilde{M}_{m-1}$), a diagonal block matrix, where $0_n$ represents a row vector with $n$ zeros and $n=3N$. The corresponding eigenvector $\vb$ of the original matrix $H$ in Eq.~(\ref{eq:As}) can be obtained by~\cite{Cao90}, 
\begin{equation}
\vb = (F_m \otimes I_n)\wb = 
\begin{bmatrix}
\uj \\
\uj\omega_m^j\\
\vdots \\
\uj\omega_m^{(m-1)j}\\
\end{bmatrix}
\label{eq:v}
\end{equation}
where $I_n$ is an identity matrix of dimension $n$. Now let $\uj$ be,
\begin{equation}
\uj = 
\begin{bmatrix}
r_1e^{i\theta_1} \\
r_2e^{i\theta_2} \\
\vdots \\
r_ne^{i\theta_n} \\
\end{bmatrix}
\end{equation}

Then, 
\begin{equation}
\vb = 
\begin{bmatrix}
r_1e^{i\theta_1} \\
\vdots \\
r_ne^{i\theta_n} \\
r_1e^{i(\theta_1+\phi)} \\
\vdots \\
r_ne^{i(\theta_n+\phi)} \\
\vdots \\
r_1e^{i(\theta_1+(m-1)\phi)} \\
\vdots \\
r_ne^{i(\theta_n + (m-1)\phi)} \\
\end{bmatrix}
=
\begin{bmatrix}
r_1cos(\theta_1) \\
\vdots \\
r_ncos(\theta_n) \\
r_1cos(\theta_1+\phi) \\
\vdots \\
r_ncos(\theta_n+\phi)\\
\vdots \\
r_1cos(\theta_1+(m-1)\phi) \\
\vdots \\
r_ncos(\theta_n + (m-1)\phi) \\
\end{bmatrix}
+ i
\begin{bmatrix}
r_1sin(\theta_1) \\
\vdots \\
r_nsin(\theta_n) \\
r_1sin(\theta_1+\phi) \\
\vdots \\
r_nsin(\theta_n+\phi)\\
\vdots \\
r_1sin(\theta_1+(m-1)\phi) \\
\vdots \\
r_nsin(\theta_n + (m-1)\phi) \\
\end{bmatrix} 
= \pb + i\qb,
\label{eq:pb}
\end{equation}
where $\phi = 2\pi j/m$.

Since H is a real symmetric matrix, it means that both $\pb$ and $\qb$ are the eigenvectors of $H$, having the same eigenvalue. Next, we will show that $\pb$
and $\qb$ are orthogonal to each other, i.e., $\pb \cdot \qb = 0$.

Consider $\pb_u$ and $\pb_d$ that are obtained by rotating $\pb$ upward or downward by $n$ elements, respectively:   
\begin{equation}
\pb_u = 
\begin{bmatrix}
r_1cos(\theta_1+\phi) \\
\vdots \\
r_ncos(\theta_n+\phi)\\
\vdots \\
r_1cos(\theta_1+(m-1)\phi) \\
\vdots \\
r_ncos(\theta_n + (m-1)\phi) \\
r_1cos(\theta_1) \\
\vdots \\
r_ncos(\theta_n) \\
\end{bmatrix},
\pb_d = 
\begin{bmatrix}
r_1cos(\theta_1+(m-1)\phi) \\
\vdots \\
r_ncos(\theta_n + (m-1)\phi) \\
r_1cos(\theta_1) \\
\vdots \\
r_ncos(\theta_n) \\
\vdots \\
r_1cos(\theta_1+(m-2)\phi) \\
\vdots \\
r_ncos(\theta_n+(m-2)\phi)\\
\end{bmatrix}
\label{eq:pbu}
\end{equation}
Now it is evident that $\pb_u = \pb  cos(\phi) - \qb  sin(\phi)$. $\pb_d = \pb  cos(\phi) + \qb  sin(\phi)$. Since $\pb_u \cdot \pb_u = \pb_d \cdot \pb_d$, we have, $\pb \cdot \qb = 0$.

Since $\pb$ and $\qb$ are both eigenvectors of $H$ and share the same eigenvalue, the normal modes that these eigenvectors represent have 
a degeneracy of 2, which is consistent with small oscillations of 
$m$ identical masses on a hoop (see section~\ref{sec:oscillation}). It is worth noting that that this kind of degeneracy originates purely from symmetry in structure and is different from the kind of degeneracy 
of normal modes caused by structure uncertainty~\cite{Na16b}.

%

The degeneracy of 2 in normal modes can be understood also in an alternative way. Notice that,
\begin{equation}
\tilde{M}_{m-j}  = 
\sum_{k=0}^{m-1} \omega_m^{k(m-j)}A_k
 = \sum^{m-1}_{k=0}\bar{\omega}_m^{kj}A_k =\overline{\tilde{M_j}}
\label{eq:cc}
\end{equation}
That is, ${M_{m-j}}$ is a complex conjugate of $M_j$. 
It is clear both $\tilde{M}_j$ and $\tilde{M}_{m-j}$ are Hermitian matrices and have real 
eigenvalues. Furthermore, since they are complex conjugate to each other, $\tilde{M}_j$ and $\tilde{M}_{m-j}$ have the same set of eigenvalues. {Therefore, the normal modes computed from $\tilde{M}_j$ and $\tilde{M}_{m-j}$ using eq.~(\ref{eq:v}) have a degeneracy of 2 in general. The only exception to this  is that $\tilde{M}_0$ (eq.~(\ref{eq:M0})), or 
$\tilde{M}_{m/2}$ (eq.~(\ref{eq:anti})) when $m$ is even, is a real matrix. 
 The modes computed from them have no degeneracy. Later, we will 
show that the modes computed from $\tilde{M}_0$ are perfectly symmetric modes, while 
modes computed from $\tilde{M}_{m/2}$ are perfectly anti-symmetric modes. }

\subsection{Representing Hessian Matrix as a Circulant Block Matrix}
For a symmetric ring structure with $m$ subunits, it has a m-fold cyclic symmetry ($C_m$). Let $\R$ represent a rotation of $2\pi/m$ degree. Representing the motions of each subunit in its local frame, 
the Hessian matrix becomes a symmetric circulant block matrix: 
\begin{equation}
H = 
\begin{bmatrix}
A_{0}     & A_{1} & \dots  & A_{m-2} & A_{m-1}  \\
A_{m-1} & A_0    & \dots  & A_{m-3}      & A_{m-2}  \\
\vdots  & \vdots &	\vdots    & \vdots  & \vdots   \\
A_{2}  &  A_3      & \dots & A_0  & A_{1}   \\
A_{1}  & A_{2} & \dots  & A_{m-1} & A_0 \\
\end{bmatrix}
\end{equation}
where $A_{i}$ represents the interaction between the current subunit and the $i_{th}$ subunit down the ring (clockwise or counter-clockwise).   


The eigenvectors of $H$ are given in Eq.~(\ref{eq:v}).
The eigenvectors in the {\em global} coordinate are:
\begin{equation}
\vb = 
\begin{bmatrix}
\uj \\
(I_n\otimes \R)\uj\omega_m^j\\
\vdots \\
(I_n\otimes \R^{n-1})\uj\omega_m^{(m-1)j}\\
\end{bmatrix}
\end{equation}

\subsection{The Symmetricity of Normal Modes}
For ring structures with $m$ identical subunits, 
we define the symmetricity of its modes as follows~\cite{Na16c}. For each mode $\mb_i$, we perform 
a rotation of $2\pi/m$ along its central axis. Let $\mb'_i$ be the mode after 
the rotation. 
The symmetricity of mode $i$ is defined as the overlap (or dot product) between $\mb_i$ and $\mb'_i$, i.e.,
\begin{equation}
symmetricity = \mb_i \cdot \mb'_i
\label{eq:sym}
\end{equation}
If a mode is {perfectly} symmetric along the central axis, the rotation has no effect and its symmetricity should be 1. That is, the
mode is invariant under rotations of  multiples  of $2\pi/m$. 

If we represent the motion of 
each subunit in its local frame, the effect of the rotation is the same as 
rotating the elements of $\mb_i$ by one subunit block ($n$ elements). 
If we let $\mb_i$ be $\pb$ (see Eq.~(\ref{eq:pb})), as $\pb$ indeed is an eigenvector of H, then $\mb'_i$ is the same as $\pb_u$ in Eq.~(\ref{eq:pbu}). 
Consequently, since $\pb\cdot\pb$ = 1 and $\pb\cdot\qb$ = 0, we have, 
\begin{equation}
symmetricity = \mb_i \cdot \mb'_i = \pb \cdot \pb_u = cos(\phi) = cos(2\pi j/m),
\label{eq:sym1}
\end{equation}  
where $j=0, 1, \cdots, n-1$. This means for symmetric ring structures with $m$
units, $1/m$ of the modes have symmetricity of cos(0) = 1, $1/m$ of the modes have symmetricity of $cos(2\pi/m)$, $1/m$ of the modes have symmetricity of $cos(4\pi/m)$, and so on.

\subsection{Construct Symmetric Complexes}
The biological assembly reported in PDB~\cite{Berman00} for symmetric ring structures generally are not exactly perfectly symmetric along the central axis. There is usually a small amount of deviation from the otherwise perfectly symmetric structure.  

To construct a perfectly symmetric structure model is simple. One can pick one subunit from 
a given ring structure and perform on it a series of $m$ rotations (clockwise or counter-clockwise) of $2\pi /m$ degrees along the central axis (where $m$ is the number of subunits) until it comes back to its original conformation and collect all the intermediate 
conformations, which, together with the subunit's initial conformation, form a
conformation of a symmetric ring structure. We term such a perfectly symmetric structure as {\em perfect structure} in the rest of the paper, as contrast to actual PDB structures that may or may not be perfectly symmetric.

{\bf Definition 1: Perfect Structure.} A perfect structure is a ring structure, constructed or actual, that has perfect axial symmetry. 

{\bf A perfect structure of p97.} p97~\cite{Ye01-Nature} is an important protein in the extended AAA (ATPases Associated with diverse cellular Activities) family. p97 is a symmetric hexamer and there are about a dozen of p97 structures deposited in PDB~\cite{Na16c}. One of them is 5ftl~\cite{Banerjee16}, a cryo-EM structure of p97 bound with ATP analogs. 5ftl is nearly exactly symmetric. We apply the above procedure and construct a perfect structure of p97 using chain A of 5ftl and name the new structure model {\em 5ftl-perfect}. The root mean square distance between 5ftl and 5ftl-perfect is 0.0042~\AA. Both 5ftl and 5ftl-perfect are used in our study

\section{Results}
The symmetricity of normal modes in ring structures describes the extent of
synchronization in motion between adjacent subunits along the ring. When applied to the
whole ring structure, it can reveal 
the extent of cooperativity~\cite{Yang09} among the subunits.
In the following, we will take a close investigation of symmetricity in 
ring structures and its implications. 

\subsection{Efficient Computations of Normal Modes of Ring Structures}
The circulant block Hessian matrix allows us to efficiently compute 
normal modes of ring structures. Particularly, Eqs.~(\ref{eq:As}) to (\ref{eq:pb}) show how 
the task of solving for  the normal modes of the whole multi-mer can be divided  by solving for the eigenvalues of a few matrices of a smaller size: $m$ times smaller to be precise, where $m$ is the number of subunits in the complex. Since solving for eigenvalues may take up to a cubic time to the size of 
the matrix, the computational 
gain can be enormous.   
This allows normal modes of ring structures be computed in a much smaller 
amount of time, using a much smaller amount of memory, thus making it possible
to extend the normal mode computations to larger ring structures. 

Moreover, in so doing, the normal modes are naturally grouped by their level of symmetricity. That is, the normal modes within each group have the same symmetricity (see the Methods section on symmetricity). 

This natural grouping can be highly beneficial when only modes with 
certain symmetricity is needed. For example, if all we care about are symmetric modes (i.e., symmetricity = 1), we only need to solve one $M$ matrix (see Methods)! The computation time will be further reduced. Indeed, for symmetric ring structures, it is likely that only symmetric or anti-symmetric (symmetricity=-1) are functionally important. For example, 
a previous study on p97~\cite{Na16c} indicated that only symmetric modes contribute to the conformation changes that also are symmetric.

\subsection{Perfectly Symmetric and Anti-Symmetric Modes} 
For symmetric ring structures with $m$ subunits, it is evident that $1/m$ of 
the modes are symmetric. For those ring structures with an even 
number of subunits, an additional $1/m$ of the modes are anti-symmetric. The anti-symmetric modes can be computed by solving Eq.~(\ref{eq:anti}). Structures with an odd number of subunits do not have anti-symmetric modes. Anti-symmetric modes represent motions where half the subunits (every other subunit around the ring) synchronize perfectly and move in the opposite direction to the other half of subunits. This pattern of motions may be important for the functions of some complexes whose two subsets of subunits alternate their roles in function, such as ATP binding and hydrolysis. 

It is helpful to realize that both symmetric and anti-symmetric of an $m$-subunit 
structure are symmetric modes of the same structure if structure is considered as an $m/2$-dimer that has $m/2$-fold axial symmetry. For example, the heat shock locus protein (HslU), which consists of 6 subunits, was found to behave as a hexamer when all six units are bound with ADP (or ATP)~\cite{Wang01a, Bochtler00}, and as a trimer of dimer when only every other subunit binds with an ADP~\cite{Bochtler00, Wang01b}.

\subsection{Symmetric Modes Form a Closed Subspace}
Symmetric modes can be obtained by solving Eq.~\ref{eq:M0}. It is evident that symmetric 
modes amount to $1/m$ of the total number of modes. All the symmetric modes take the form of:
\begin{equation}
{\bf S_j} = 
\begin{bmatrix}
\uj \\
\uj\\
\vdots \\
\uj \\
\end{bmatrix}
\end{equation} 
where $\uj$ is an eigenvector of $\tilde{M}_0$ in Eq.~(\ref{eq:M0}) and 
$j=0, 1, \cdots, 3N$, where $N$ is number of atoms in each subunit. 
Therefore, it is obvious that all the symmetric modes form a closed subspace in the sense that any symmetric conformation displacement can be written as a linear combination of {\bf $S_j$}'s. That is, for a symmetric conformational displacement {\bf ds}, 
\begin{equation}
{\bf ds} = 
\begin{bmatrix}
\db \\
\db\\
\vdots \\
\db \\
\end{bmatrix}
\end{equation}
we can always have: 
\begin{equation}
{\bf ds} = \sum_{i=1}^{3N} c_i {\bf S_i}
\end{equation} 
This is evident since the displacement of each unit, $\db$, in general can always be written 
as a linear combination of $\uj$'s. 
\begin{equation}
\db = \sum_{i=1}^{3N} c_i \uj.
\end{equation} 

There are a couple of important implications. First, if we are interested only in symmetric 
conformation displacements or conformation transitions, only symmetric modes are needed and 
they alone provide complete information regarding how the transition may take place. 
A second implication is that non-symmetric modes cannot be linearly combined to give a symmetric displacement.  

As we will see later,
for ring structures that are not completely symmetric,  symmetricity is partially broken due to degeneracy~\cite{Na16b} 
and the number of modes with  
perfect symmetry (i.e., symmetricity=1) is reduced
and their fraction is less than $1/m$ and consequently they no longer form 
a closed subspace. However, symmetric modes that are functional are generally of low frequency and are robust to degeneracy and can remain to be symmetric. These symmetric modes, though no longer forming a closed subspace, may still be sufficient to interpret any symmetric function-related conformation changes~\cite{Na16c}. 

\subsection{The Symmetricity of Normal Modes}
The level of symmetricity reveals the extent to which adjacent subunits along 
the ring are moving together, or the extent of their similarity.  
The theoretical prediction on symmetricity as given in Eq.~\ref{eq:sym1} states that 
the whole set of modes can be divided into $m$ groups of the same size and the modes in each 
group have the same symmetricity. The symmetricity for group $j$ ($0\le j \le m-1$) is 
$\cos(\frac{2\pi j}{m})$. The angle $\frac{2\pi j}{m}$ represents the amount of phase shift 
 between adjacent subunits. 

\begin{figure}[H]

\begin{minipage}[b]{.48\textwidth}
\centerline{\includegraphics[width=\textwidth]{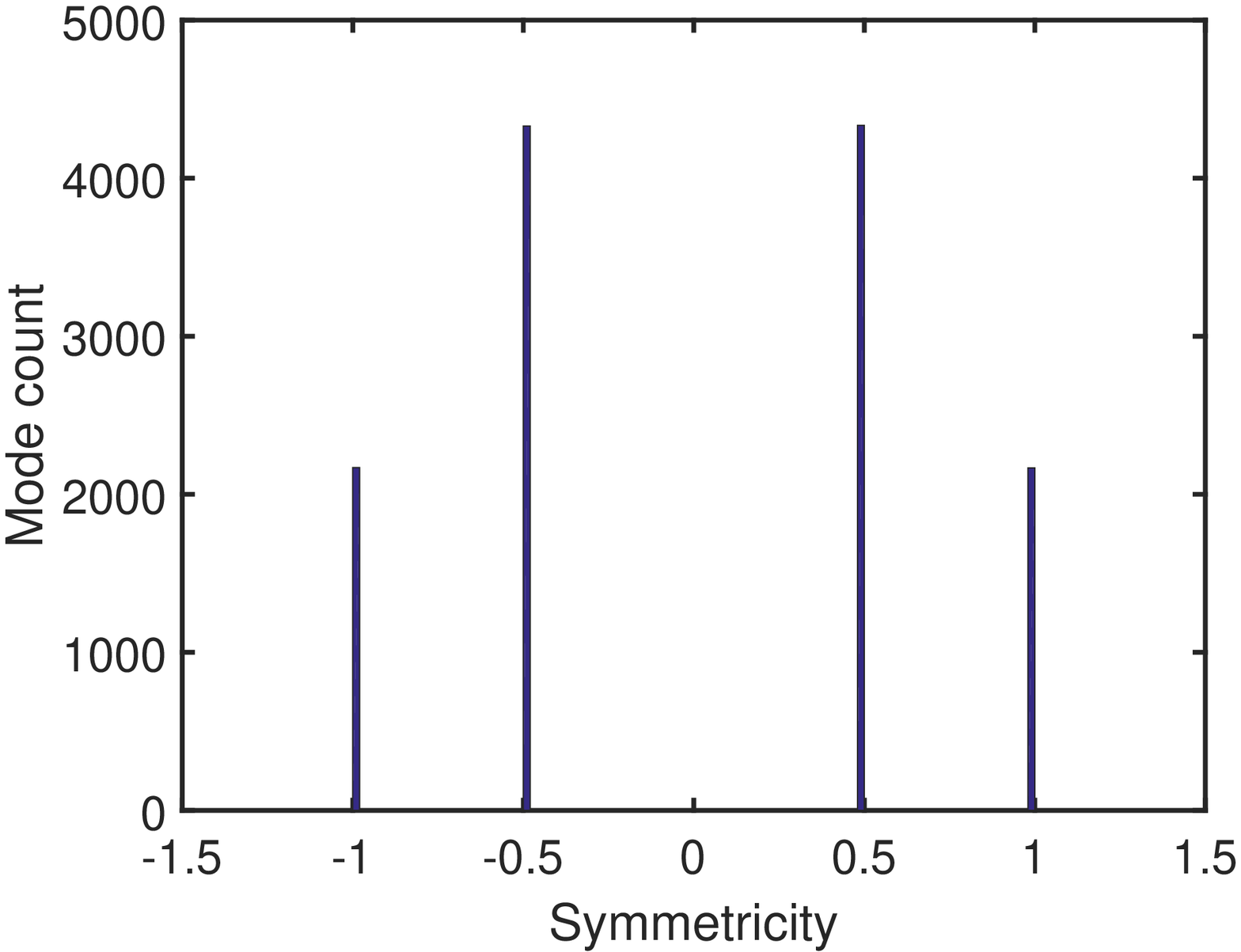}}
\centerline{(A)}
\end{minipage}

\begin{minipage}[b]{.48\textwidth}
\centerline{\includegraphics[width=\textwidth]{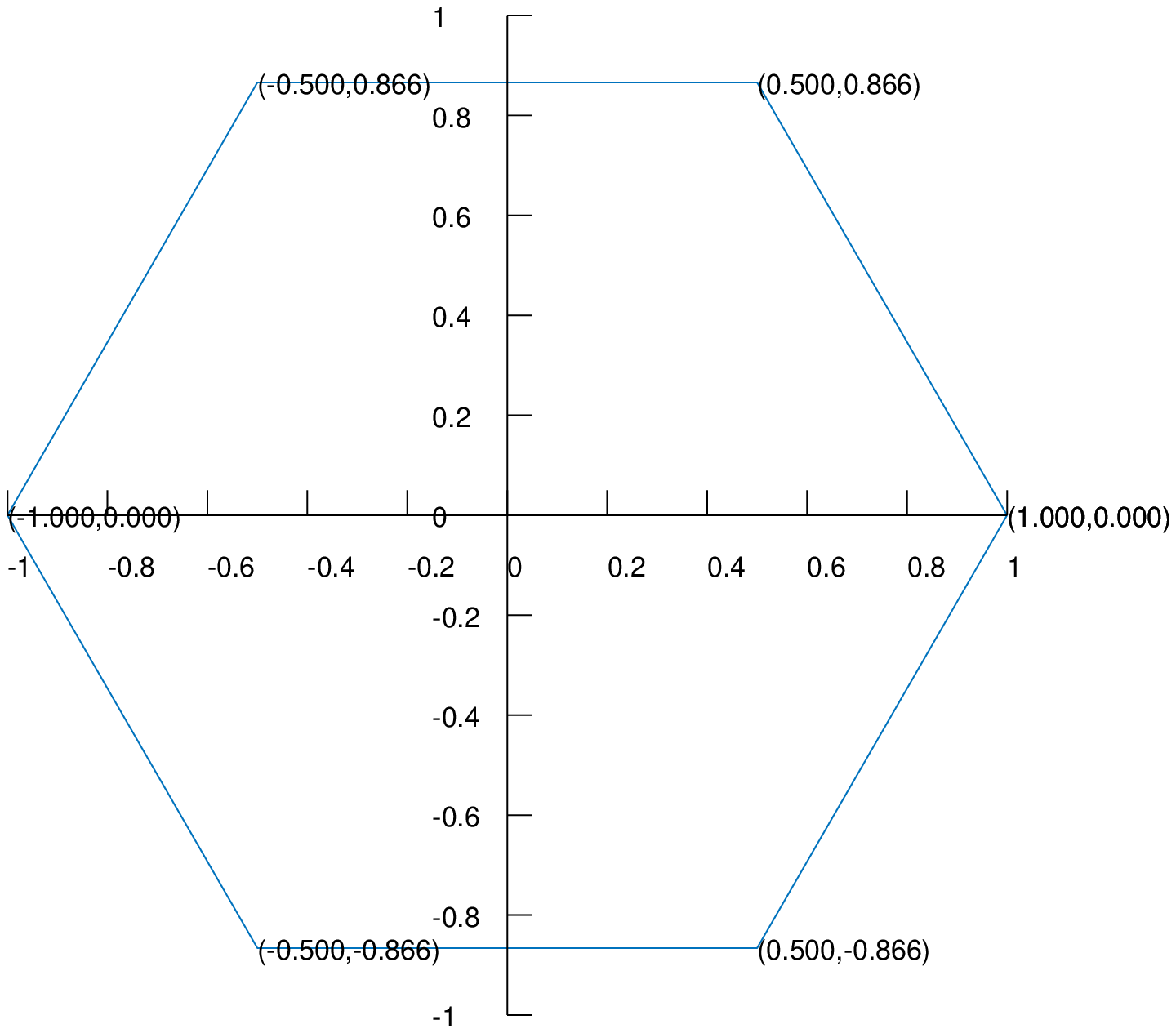}}
\centerline{(B)}
\end{minipage}

\caption{\small
}
\label{fig:hex}
\end{figure}

Figure~\ref{fig:hex}(A) shows the symmetricity distribution of the normal modes of p97, computed from a perfectly symmetric model of p97: 5ftl-perfect, which is  constructed from PDB~\cite{Berman00} structure 5ftl (see Methods). 
The computed distribution matches perfectly with the theoretical 
prediction: 1/6 of the modes have symmetricity of 1 or -1, and 1/3 of the modes have symmetricity of 0.5 (which equals to $\cos(\frac{2\pi}{6})$ or $\cos(\frac{5*2\pi}{6})$) or -0.5 (which equals to $\cos(\frac{2*2\pi}{6})$ or $\cos(\frac{4*2\pi}{6})$).

Now consider a unit polygon with $m$ 
sides as shown in Figure~\ref{fig:hex}(B),
with vertices starting at (1, 0) and proceeding counterclockwise
at angles in multiples of $2\pi/m$,
the symmetricity of each group of normal modes mentioned above is the same
as the x-coordinate values of the vertices of the polygon. Groups $i$ and $m-i$ clearly have the same symmetricity. When $n$ is even,
there is a group with symmetricity of -1, or anti-symmetric.

\begin{figure}[H]
\centerline{\includegraphics[width=0.6\textwidth]{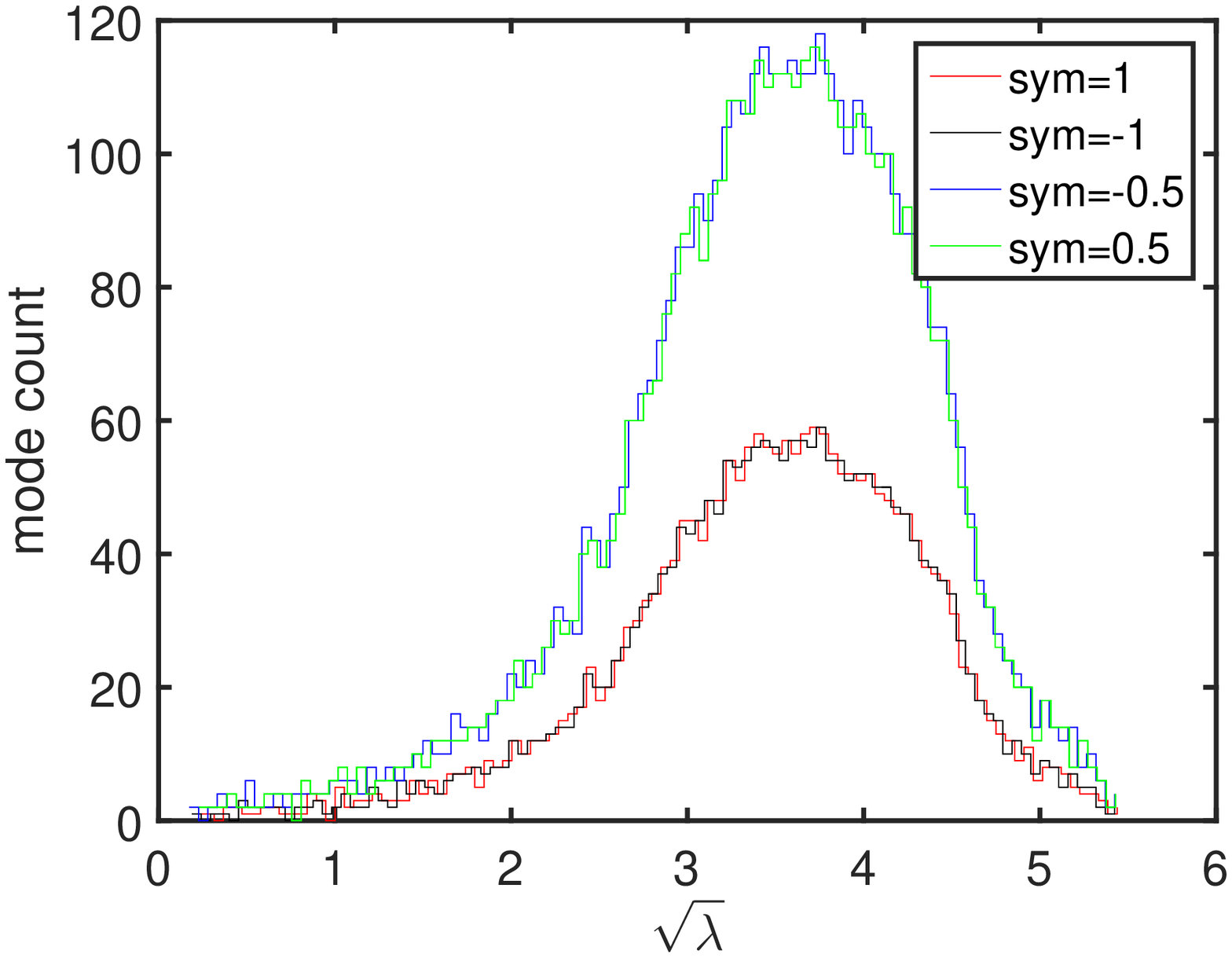}}
\caption{\small
}
\label{fig:dist}
\end{figure}

{\bf Symmetric modes distribute through the whole frequency range. }
Since Symmetric modes are obtained by solving the Hessian matrix in Eq.~\ref{eq:M0},
it is perceivable that the symmetric modes should distribute through the whole frequency range. 
Figure~\ref{fig:dist} shows the density of state distributions of modes with different
symmetricity. It is seen that 
modes with different symmetricity spread evenly and have nearly the same distribution (ignoring the scaling factor).

\subsection{Symmetricity Pattern is Partially Broken in Real Structures}
Crystal structures reported in PDB contains a set of coordinates solved 
from the structure factors. These coordinates represent structural 
data in the asymmetric units. They may not represent the complete 
complex or may contain multiple copies of the same molecule. To 
construct the complete biological assembly of a molecule based 
on the given coordinate data from the asymmetric units (as well as information 
on space group and unit cell), software such 
as PISA~\cite{pisa} or PQS~\cite{pqs} is often used. 
There are cases where the asymmetric unit contains the 
whole multimer, such as the hexamer structure of HslU (PDB-id: 1DO2~\cite{Bochtler00}) 
or p97 (PDB-id:5C18~\cite{Hanzelmann16}).

The recent advancement of cryo-EM technology makes possible the determination 
of many large structure complexes at near-atomic resolution~\cite{Bai15}. Cryo-EM usually 
assumes structure symmetry (for complexes that are symmetric) and is able to produce a whole structure assembly.

For most of the homomer structures reported in PDB, the subunits are not perfectly symmetric. There exist some slight structure deviations from the 
otherwise perfectly symmetric structures. The effective degeneracy of normal
modes~\cite{Na16b} dictates that modes are degenerate under slight structure variations and may mix together with
other modes with similar frequencies.
Consequently, not all the modes computed from a perfectly symmetric structure will maintain their symmetricity in reality. 

%

\begin{figure}[H]
\centerline{\includegraphics[width=0.6\textwidth]{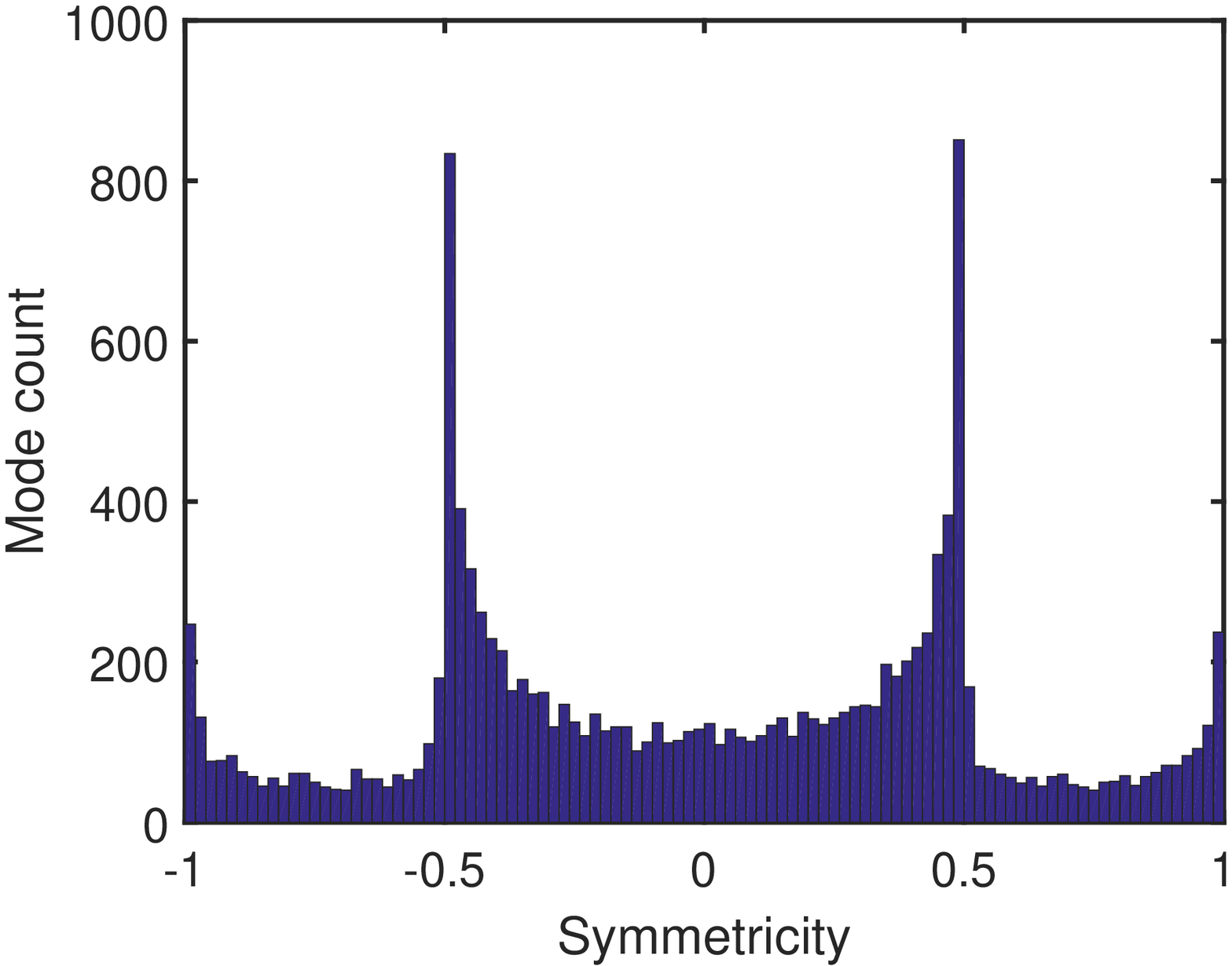}}
\caption{\small
}
\label{fig:universal}
\end{figure}

Figure~\ref{fig:universal} shows the symmetricity plot of p97 (pdb-id: 5ftl) as computed by ANM~\cite{Atilgan01}. The result is the same as that in Figure~\ref{fig:hex}(A) except that the original PDB structure of 5ftl is used here. 
This cryo-EM structure of p97 (pdb-id: 5ftl) as reported in PDB is nearly perfectly axially symmetric. The root mean square deviation between this structure and the one constructed above by selecting chain $A$ of 5ftl and rotating multiples of 60 degrees is only 0.0042~\AA.

From the figure it is seen that, comparing to the 2,000+ symmetric modes (i.e., symmetricity=1) of the perfectly symmetric structure in Figure~\ref{fig:hex}(A), there are only about 200 modes in the actual structure that have a symmetricity of nearly 1 ($\ge$ 0.98). 

\begin{figure}[H]
\centerline{\includegraphics[width=0.6\textwidth]{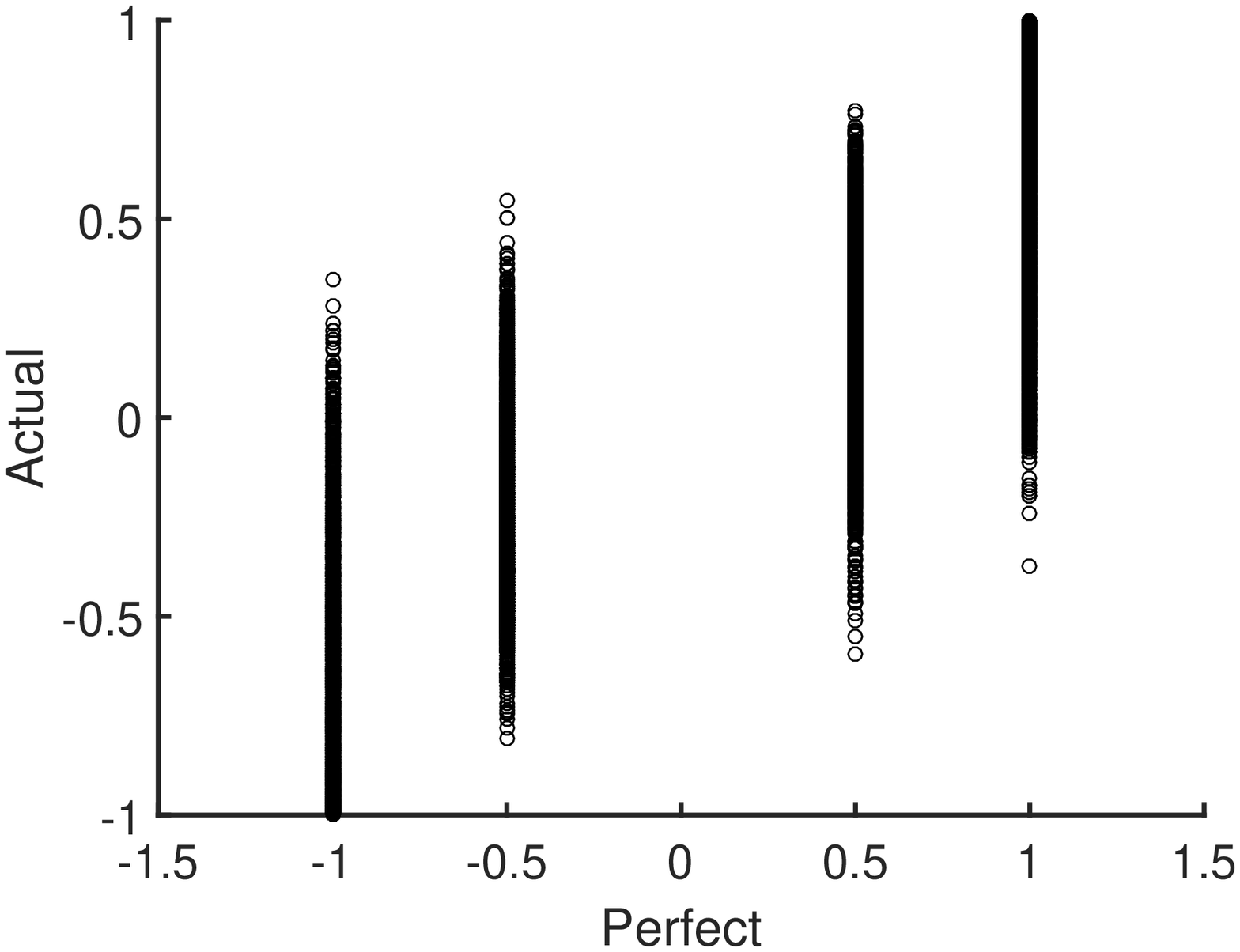}}
\caption{\small
}
\label{fig:actual}
\end{figure}

What are these 200 modes? Are they just a subset of the 2000+ symmetric modes of the perfectly symmetric structure? 
To answer this question, we find, for each mode of the perfect structure, the best matching mode in the modes of the actual structure.  
Figure~\ref{fig:actual} gives a scatter plot of 
the symmetricity of the modes of the perfect structure and the symmetricity of their best matching modes in the actual structure. The figure shows that the  
modes with high symmetricity ($\ge$ 0.8) of the actual structure all match to modes with 
symmetricity of 1 in the perfect structure. The symmetric modes (symmetricity=1) of the perfect 
structure match to the modes of the actual structure with a wide-spread range of symmetricity. 
Of all the symmetric modes of the perfect structure, only a small percentage of them 
are able to preserve their symmetricity.

\begin{figure}[H]
\centerline{\includegraphics[width=0.6\textwidth]{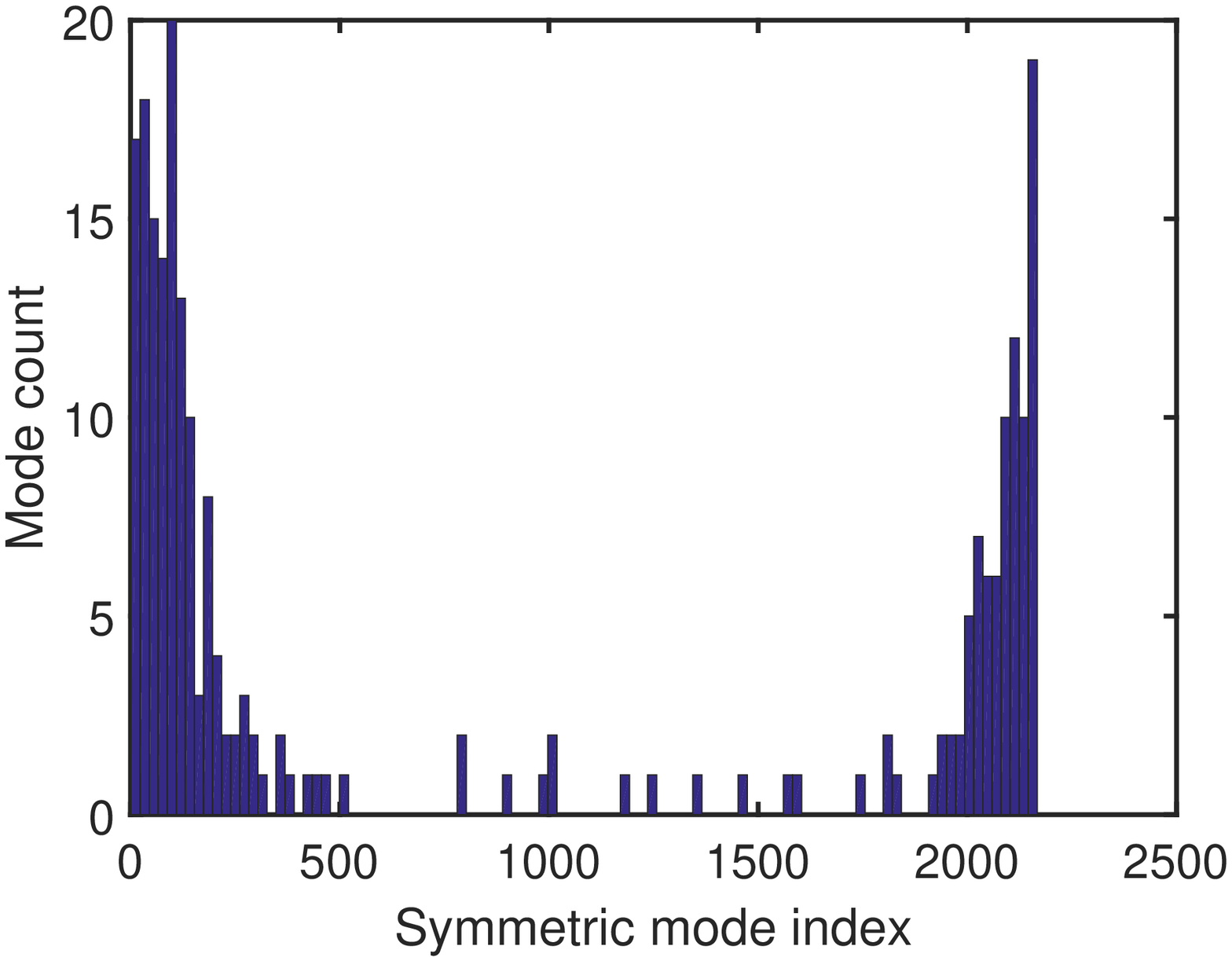}}
\caption{\small
}
\label{fig:degeneracy}
\end{figure}

The next question is, what are these modes that are able to preserve their perfect symmetricity of 1 and why? What happens to the other symmetric modes?
Figure~\ref{fig:degeneracy} shows the frequency distribution of the modes that are able to preserve their perfect symmetricity of 1. Most of these modes fall into either the low frequency end (low mode indices) or the high frequency end (high mode indices).  

\begin{figure}[H]
\centerline{\includegraphics[width=0.5\textwidth]{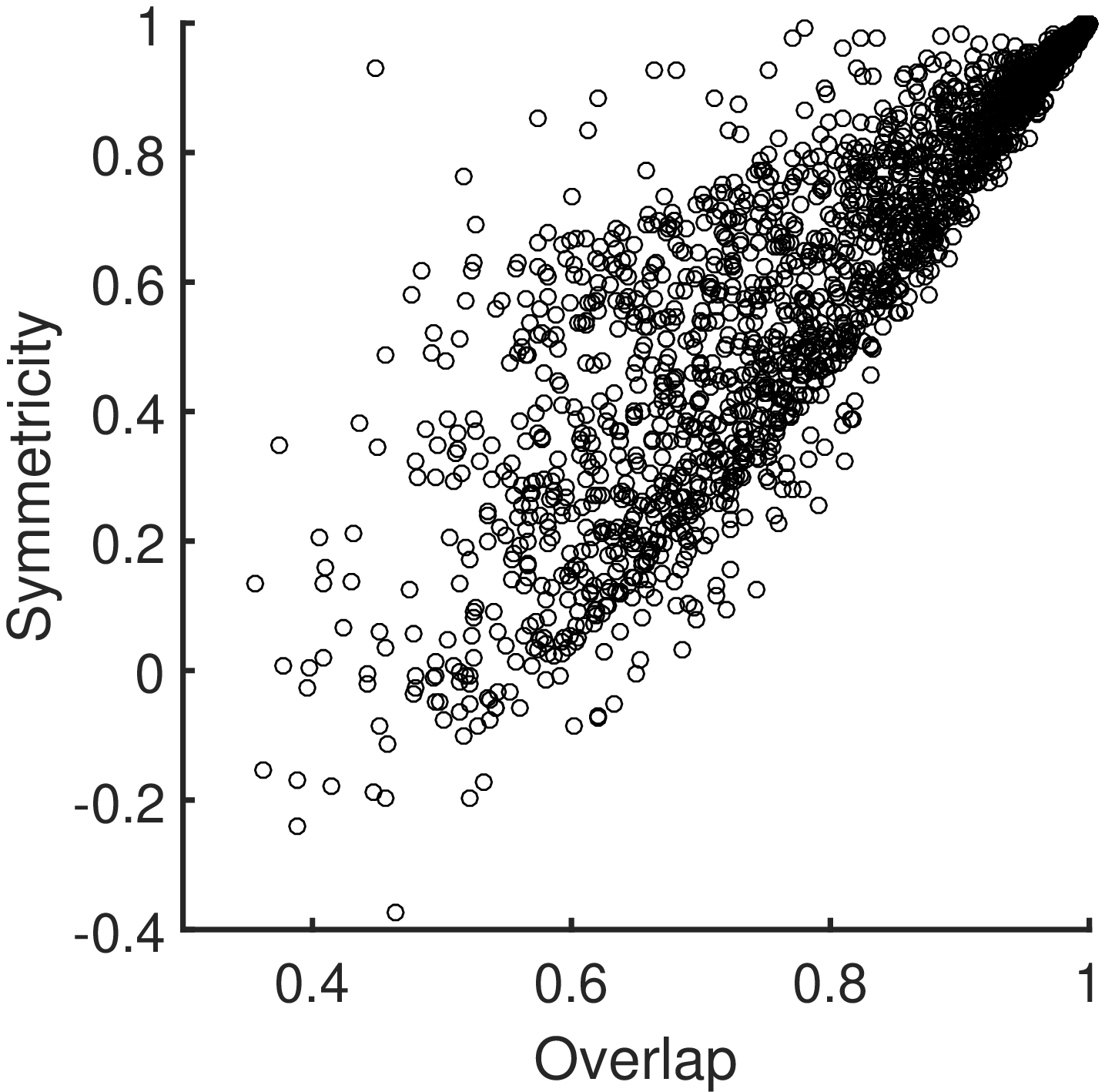}}
\caption{\small
}
\label{fig:overlap}
\end{figure}

To find out why the modes at either the low frequency end or the high
frequency end can preserve their symmetricity and 
what happens to the other symmetric modes, 
we plot in Figure~\ref{fig:overlap}, for the 2,167 symmetric modes of the perfect
structure, 
the symmetricity of their besting matching modes in the actual structure and the  overlaps between them and their best matching modes. 
Not surprisingly, a strong correlation is found: 
the modes that are mostly preserved (or unchanged) under 
structure variation, as indicted by a large overlap, also preserve most of their symmetricity.
Thus, to the question raised earlier, ``Why are the modes at the low
or high frequency end able to preserve their symmetricity?'' the answer is that these modes are robust to small structural changes and remain mostly unchanged (with a large overlap). 
This is thus consistent with our 
previous finding that modes at the low or high frequency end are
less degenerate~\cite{Na16b}. 
On the other hand, the low overlaps of the other modes indict that
they have deformed greatly under the structure deviation. 
Figure~\ref{fig:overlap} shows that 
non-degenerate modes are the ones that can maintain their symmetricity, 
while degenerate modes generally cannot. 

It is perceivable that for some systems 
symmetric modes are critical to the realization of functions. The ability
to  preserve their symmetricity is consequently important. 

\subsection{Obtain Symmetric and/or Anti-Symmetric Modes from Nearly-Symmetric Structures}

Practically speaking, 
most ring structures deposited in PDB do not have exact symmetry. The coordinates of the atoms in each subunit may be slightly off from their 
otherwise perfectly symmetric locations. Moreover, we do not 
expect symmetric ring structures such as p97 to maintain a mathematically exact symmetry while they function in cell. 
For such structure models that have nearly exact symmetry, how do we obtain symmetric modes efficiently? Apparently, we cannot apply Eq.~(\ref{eq:M0}) or (\ref{eq:anti}) if the structure is not 
exactly symmetric. 

One possible solution is to choose not to take advantage of the symmetry and compute the normal modes using the whole structure. The drawback is that this can become computationally too costly, especially for large systems.
Another possible solution is to reconstruct a perfectly symmetric 
structure from one of the subunits by applying axial symmetric rotations and then apply Eq.~(\ref{eq:M0}) to compute modes. The drawback of this approach is that the modes computed by the reconstructed structure (which has 
a perfect symmetry) are generally somewhat different from those computed with the original (PDB) structure. 
Such difference reflects the degeneracy of protein normal modes~\cite{Na16b}. 

A more ideal solution is to be able to obtain symmetric modes that are robust to the structure differences among the subunits. 
Is there an efficient way to obtain the subset of symmetric modes that are robust to structure variations without solving the Hessian matrix of the whole multi-mer? 
To answer this question, it is helpful to realize that the subunits are highly
similar to one another in structure and they only deviate slightly from the otherwise perfectly symmetric structure. 
In our previous work, we have shown that some normal modes, especially 
those at low frequency end, are robust to small structure deviations while others are not and become degenerate~\cite{Na16b}.  For these non-degenerate modes, 
the slight deviation from the perfect symmetry does not disturb them. 

For this reason, we determine symmetric modes that are robust to 
small structure deviations in the following way. 
First, we use each subunit in turn as the center and use Eq.~(\ref{eq:M0}) compute the symmetric modes. We then find the symmetric modes that are common to all subunits (modes are considered common or the same if the overlaps between them are greater than 0.99). These modes are robust to structure 
differences among the subunits. 

\begin{figure}[H]
\centerline{\includegraphics[width=0.6\textwidth]{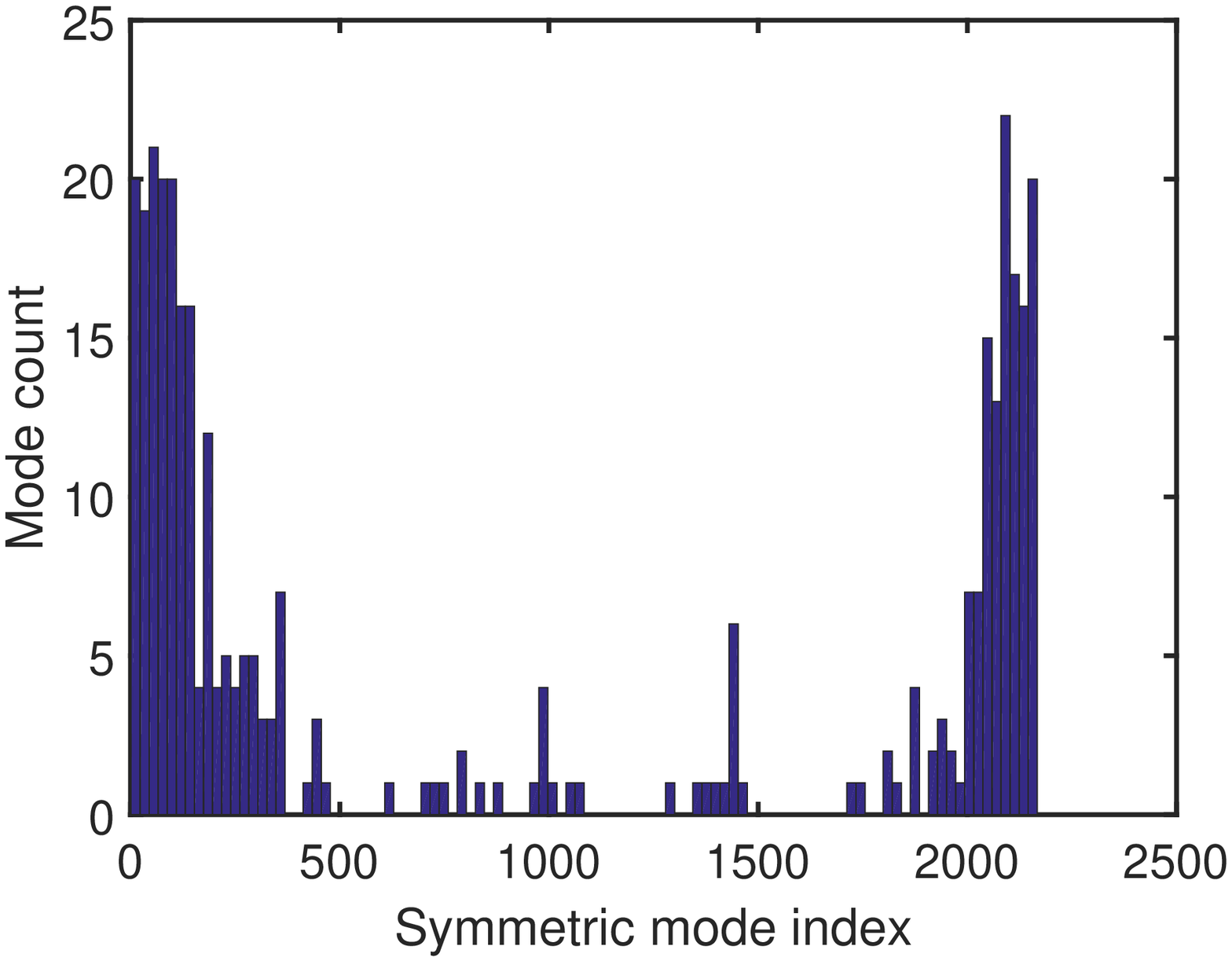}}
\caption{\small
}
\label{fig:common}
\end{figure}

Figure~\ref{fig:common} shows the index/frequency distribution of the symmetric modes that are common to all 
subunits. 
The modes computed from each subunit are compared with
the 2,167 symmetric modes of a perfectly symmetric structure of p97: 5ftl-perfect. Modes that are common to all 
subunits are then shown in Figure~\ref{fig:common} as a histogram.  
The figure shows that the modes that are common to all subunits 
consist mostly of the modes at the low frequency end, representing domain motions, or modes at the high frequency end, representing localized motions. The distribution in Figure~\ref{fig:common} is 
similar to that in Figure~\ref{fig:degeneracy}, which is computed using the whole structure. Indeed, Of the  
first 50 lowest frequency symmetric modes of the actual structure, 
which are more likely to 
be functionally important than the other modes, 47 are included 
among the symmetric modes 
common to all subunits as identified above.  

There is also a drawback with this last approach. The set of symmetric modes that are common to all subunits are not exactly the same as the set of 
symmetric modes computed from the whole complex, as indicted above. However, it is probable that both sets capture all the symmetric modes that are functionally important and thus their small difference is not a problem. When this is in doubt, one may also compute the modes from the whole complex (given that it is computationally feasible) and compare the two sets of modes closely.

\section{Discussion}

In this work, we have looked into the symmetricity of protein normal 
modes and its implications. Symmetricity is introduced to define how similar the motions of the adjacent subunits are in symmetric structures. 
We show that the symmetry in structure can be taken advantage of to compute the normal modes of symmetric ring structures most efficiently. We then
present a new theoretical result on the symmetricity of the normal modes of ring 
structures and confirm the theoretical prediction with computational results. 
Lastly, we show that the symmetricity pattern is broken in real structures 
due to the degeneracy of protein normal modes under structure 
variations. The work has several important implications.

{\bf The importance of symmetric and anti-symmetric modes in biological functions.} 
It is perceivable that some functional 
motions of structurally symmetric systems are symmetric. The present work
shows how to obtain symmetric normal modes separately in a fraction of the time that is 
otherwise needed to get all the modes. Anti-symmetric modes also are likely to be functionally important, as they represent a
motion pattern in which every other subunit in a ring structure synchronizes  perfectly and moves in the opposite direction to the other half of the subunits. As symmetric modes, anti-symmetric modes also can be 
obtained separately, using Eq.~(\ref{eq:anti}).  
The heat shock protein (HslU)~\cite{Bochtler00}, for example, which also is a hexamer, was found 
to allosterically bind ADP at every other subunit~\cite{Bochtler00,Wang01b}. 
The behavior may be best understood using anti-symmetric motions, where every other subunit
has perfectly synchronized motions.
It is likely that most AAA+ proteins~\cite{Hanson05} that employ a threading mechanism may use symmetric 
motions along the central pore as their primary function motions. 
A deeper understanding of the motion patterns of 
their symmetric modes may help us better understand, for example, how 
protein unfolding and degradation are carried out.

On the other hand, it is possible that some molecular 
systems, though having symmetric structures, may not  utilize symmetric motions as their primary functional movements. This is probably the case with GroEL/GroES
complex~\cite{Xu97}. A major function of GroEL/ES is to provide a conducive environment for 
proteins to unfold and refold. It may rely on different kinds of 
motions such as twisting or stretching  to accomplish this purpose. GroEL/GroES 
is a heptamer with seven subunits, which is a primer number. It cannot function as a trimer of dimers or a dimer of trimers as the hexmatic HslU can. This might be 
the reason why its function motions are less or even not symmetric. It is possible that multi-mers with an even (or a highly composite) number  of subunits should have more symmetric functional motions than those with an odd (or even a prime) number of subunits~\cite{Kidera12}.

Conformation changes that are symmetric require symmetric modes.
It is likely that
in some complexes symmetric motions are critical to the functions, such as threading. Consequently, the system must have a way to preserve at least some of its symmetric modes, especially those are key to function. 
No all symmetric modes are robust to structure variations: some modes are more robust to structure perturbations than the others. The former preserve their motion patterns under small structural variations while the latter are unraveled under the same condition.
Understanding the structural reason why 
some symmetric modes are preserved while the others are not should be useful.  
We show that the preserved symmetric modes are those with
no degeneracy~\cite{Na16b},  mostly at either the low frequency end 
or the high frequency end. 
Non-degenerate normal modes are more likely to be functional~\cite{Na16b}.

{\bf The importance of symmetry in structure to symmetric motions.} 
Identical protein units of some types can come together to form homoligomers. Whenever this is feasible, the interactions among them often drive them naturally to 
form beautiful complexes of certain symmetry. The symmetric structure is
energetically favored, corresponding to at least a stable local minimum in 
the energy landscape. Consequently, structure deviations from the symmetry are understandably disfavored. On the other hand, our work here shows that 
there is probably a dynamical 
reason why symmetry in structure ought to be maintained. Figure~\ref{fig:universal} shows that a structural deviation as small as 0.0042\AA~ can significantly affects the symmetricity distribution of the normal modes. We find  that scarcely any symmetric modes are left (data not shown) when structure deviation is greater than 0.1~\AA~for p97. This means that if symmetric motions are critical to a complex's function, then maintaining the symmetry in structure (i.e., not having a significant deviation from it) not only makes the structure look appealing
but also is important to preserving its key motion patterns and thus its functions as well. Symmetry in nature is not only aesthetically pleasing but may exist also for survival.

{\bf Degeneracy.} For a circularly symmetric structure with $m$ subunits, i.e., with C$_m$ symmetry, what should be the degree of the degeneracy of its normal modes? Since the structure is $m$-fold 
symmetric, does it mean its normal modes have a degeneracy of $m$? Group theory and our work here show that for symmetric or anti-symmetric modes (whose symmetricity equals to 1 or -1), there is no degeneracy. For all the other modes, the degeneracy is 2. 
For any pair of degenerate modes $\pb_1$ and $\pb_2$, it is helpful to know 
that any linear combination of $\pb_1$ and $\pb_2$ also represents a valid 
mode~\cite{Na16b}.
Besides the degeneracy originating from symmetry, there exists also degeneracy due to small structure deviations~\cite{Na16b}. This latter kind of degeneracy can cause some of symmetric modes to become degenerate with other modes, thus losing their perfect symmetry. However, functionally important symmetric modes should somehow be able to preserve their symmetricity.

{\bf C$_2$ symmetry.}
Having only two subunits (a dimer), structures with $C_2$ symmetry is a special case of the general 
circularly symmetric structures with $m$ subunits ($C_m$). 
First, structures with $C_2$ symmetry have only 
symmetric (symmetricity=1) and anti-symmetric (symmetricity=-1) modes. This means none of their 
modes are degenerate! In all the modes, the motions  of two subunits are either fully symmetric, or fully anti-symmetric. Being the simplest case of structures with cyclic symmetry, $C_2$ structures may be ideal candidates for investigating how the symmetry in normal modes may be broken when the symmetry in structure is broken under small structure deviations. It will be helpful to 
%
quantitatively characterize symmetric modes that are robust to 
structure perturbations and 
to understand why small structure perturbations do not alter them. 

{\bf Extension to other symmetric structures.}
Here we focus only on symmetric ring structures of $C_n$ symmetry. 
It would be interesting to know what normal modes are like for structures with other kinds of symmetry (such as dihedral symmetry) 
and how their normal mode patterns are related to function. 
There are many symmetric complexes and they 
employ a number of different kinds of symmetry.
Why do proteins form such symmetric complexes? One plausible reason given is that interactions drive it. Symmetric homomers are favored since interactions between the same structures are more favored than those between different structures~\cite{Shakhnovich07}. Another reason for the existence of symmetric complexes is for structure purpose. Symmetric complexes are formed, for example, to create a 
channel for protein unfolding and degradation~\cite{Hanson05}, or to create a closed chamber for protein unfolding and refolding~\cite{Xu97}, etc. Indeed, the existence of symmetry in structures in nature has 
inspired scientists to bio-engineer
new structures with desired symmetry and geometry  
using  nucleic acids~\cite{Seeman10} or proteins~\cite{LaiYeates12,KingBaker12}. 
Yet another reason could be for  
functional purpose:  the structure and the symmetry
in the structure are needed for achieving certain patterns of motions that are functionally critical. Regarding this no much is known. Future studies in this area could reveal interesting insights and offer inspirations.  
It is foreseeable that symmetry-based ideas may be employed also 
to design structures with desired motion patterns someday 
in the future. 

{\bf Link to small oscillations of $n$ identical masses on a frictionless hoop.}
As pointed out earlier in the paper, there is a tight link between symmetric ring structures and 
a system of $n$ identical masses connected with identical springs on a frictionless hoop, which is a commonly used an example on small oscillations in classical mechanics~\cite{Goldstein50}. If we model each subunit as a sphere, a ring structure will become such a system of $n$ masses. It is known that such an oscillating system has a zero mode and the rest of the modes (except for one if $n$ is even) have a degeneracy of 2, just as we have shown for symmetric ring structures.  

\section*{Acknowledgment}
Funding from National Science Foundation (CAREER award, CCF-0953517) is gratefully acknowledged.


\section*{}
\pagebreak
\section*{Figure Legends}

\subsection*{Figure 1.}
(A) The symmetricity distribution of the normal modes of a perfectly symmetric hexemer (p97). The structure is constructed manually using subunit A of PDB structure 5ftl (see the Methods section). (B) The symmetricity values shown in (A) are the same as the abscissa values of the vertices of a unit hexagon.

\subsection*{Figure 2.}
 Symmetric modes distribute through the whole frequency range. Frequency values are obtained by simply taking the square root of the eigenvalues ($\lambda$'s).  The ANM model~\cite{Atilgan01} is used, whose cutoff distance is set at 13~\AA~ and whose 
 spring constant is set at 1. 100 bins are used. 

\subsection*{Figure 3.}
The symmetricity of an actual structure (5ftl) that is nearly perfectly symmetric. A uniform bin size of 0.02 is used and there are 100 bins. The RMSD between this structure and the manually constructed perfect structure (5ftl-perfect) is only 0.0042~\AA. However, the symmtricity distribution is significantly affected and becomes more widely spread than that in Figure~\ref{fig:hex}(A), though the peaks remain
clearly identifiable. In total, only 237 modes maintain a very high symmetricity (greater than 0.98, in the rightmost bin). In comparison, 2,167 modes, or one sixth of the total number of the modes, have 
perfect symmetricity (equal to 1) 
in Figure~\ref{fig:hex}(A).   

\subsection*{Figure 4.}
The symmetricity of the modes of a perfect structure (5ftl-perfect) and that of their best matching modes in the actual structure (5ftl). 

\subsection*{Figure 5.}
The indices of the 237 symmetric modes that preserve their symmetricity under 
small structure variations. 
Most of these modes fall into the low frequency or the high frequency end.

\subsection*{Figure 6.}
Between the 2,167 symmetric modes of the perfect structure (5ftl-perfect) and their best matching modes in the actual structure (5ftl),  
modes that maintain a high overlap maintain also a high symmetricity in 
the actual structure (see the text).

\subsection*{Figure 7.}
Symmetric modes that are common to all subunits of a p97 structure (pdb-id: 5ftl).
Symmetric modes are computed using Eq.~(\ref{eq:M0}) as each subunit in turn is used as the center. The modes computed from each subunit are then compared with the symmetric modes of a perfect structure of p97: 5ftl-perfect. The indices of the symmetric modes that are common to all subunits fall mostly into either the low frequency or the high frequency end.

\pagebreak

\section*{}

\newpage

\newpage

\newpage

\newpage

\newpage

\newpage

\end{document}